\documentclass[graybox, nosecnum]{svmult}

\usepackage{mathptmx}       
\usepackage{helvet}         
\usepackage{courier}        
\usepackage{type1cm}        
\usepackage{makeidx}         
\usepackage{graphicx}        
\usepackage{multicol}        
\usepackage[bottom]{footmisc}
\usepackage{hyperref}        
\usepackage{soul}            
\hypersetup{colorlinks=true,urlcolor=blue}
\usepackage[square,numbers]{natbib}

\usepackage{amsmath,amssymb,tablefootnote}
\usepackage{enumerate}

\makeindex             

\begin{document}
\title*{Distortion of Gravitational-wave Signals by Astrophysical Environments}
\author{Xian Chen  \thanks{corresponding author}}
\institute{Xian Chen \at Astronomy Department, Peking University, 
Yi He Yuan Lu 5, Beijing 100871, P. R. China, 
\at Kavli Institute for Astronomy and Astrophysics at Peking University,
Yi He Yuan Lu 5, Beijing 100871, P. R. China 
\at \email{xian.chen@pku.edu.cn}
}
%
%
\maketitle
\abstract{Many objects discovered by LIGO and Virgo are peculiar because they
fall in a mass range which in the past was considered unpopulated by compact
objects.  Given the significance of the astrophysical implications, it is
important to first understand how their masses are measured from
gravitational-wave signals.  How accurate is the measurement?  Are there
elements missing in our current model which may result in a bias?  This chapter
is dedicated to these questions. In particular, we will highlight several
astrophysical factors which are not included in the standard model of GW
sources but could result in a significant bias in the estimation of the mass.
These factors include strong gravitational lensing, a relative motion of the
source, a nearby massive object, and a gaseous background.  } 

\section{Keywords} 

Gravitational waves, Chirp mass, Redshift, Acceleration, Hydrodynamical friction. 

\section{Introduction}

The detection of gravitational waves (GWs) by LIGO and Virgo has reshaped our
understanding of stellar-mass black holes (BHs). Before the detection,
stellar-mass BHs were exclusively found in X-ray binaries. Their masses fall in
a range of about $5-20\,M_\odot$.  Below or above this range, no (stellar-mass)
BH was detected. There were also speculations based on theoretical grounds that
stellar evolution can hardly produce a BH between $3$ and $\sim5\,M_\odot$
(known as the ``lower mass gap''\index{mass gap}) or between $50$ and $\sim120\,M_\odot$ (the
``upper mass gap'').  Nevertheless, the very first GW source detected by LIGO,
GW150914, had set a new record for the mass of BHs: GW150914 is formed by two
BHs each of which is about $30\,M_\odot$ \cite{ligo16a}.  Later, during the
first and second observing runs, LIGO and Virgo had detected ten mergers of
binary black holes (BBHs).  The majority of them contained BHs heavier than
$20\,M_\odot$ prior to the coalescence \cite{abbott19population}.  More
recently, LIGO and Virgo have detected a merger product of $3.4\,M_\odot$,
clearly inside the lower mass gap \cite{abbott20GW190425}, as well as a merger
of two BHs both above $60\,M_\odot$, well inside the upper mass gap
\cite{abbott20GW190521}.

The masses of these newly found BHs deserve an explanation.  While many
astrophysicists believe that our model of BH formation and evolution needs an
modification, a few are pondering over a rather different question. Is
it possible that somehow we have measured the masses inappropriately? This is a
valid question because mass is not a direct observable in GW astronomy. The
observable is a wave signal, which can be characterized by a frequency $f$ and
an amplitude $h$, both could evolve with time.  To derive from these observables
a mass for the source, a model is needed. Needless to say, different models
could yield different masses.

What is the current standard model?  Take BBHs for example.  The running
assumption is that the orbital dynamics is determined only by GW radiation and
the detected waveform is different from the emitted one only by a cosmological
redshift\index{redshift}.  More specifically, this assumption implies that the BBH is isolated
from other astrophysical objects or matter, and is not moving relative to the
observer.  Under these circumstances, the orbit of the binary shrinks according
to the GW power predicted by the theory of general relativity. The waveform is
also straightforward to calculate. It is dictated mainly by the acceleration of
the mass quadrupole moment. Given the simplicity of the physical picture of a
merging BBH, many people believe that the precision in the measurement of the
mass is limited not by our theory, but by the observation.

However, the universe is more sophisticated than what the standard model
depicts.  First, all astrophysical objects are moving relative to us, including
BBHs. The motion induces a Doppler shift to the frequency of the GWs, but the
frequency, as is mentioned earlier, is an observable essential to the
measurement of the mass.  Second, BBHs do not necessarily form in isolation.
They could form, for example, in triple systems (e.g. \cite{miller02hamilton}
and references therein).  The interaction with the tertiary may alter the
orbital dynamics of a BBH, and hence affects our interpretation of the GW
signal. Note that the tertiary could be a supermassive BH (SMBH) of more than
$10^6\,M_\odot$, according to several recent theoretical studies (pointed out
by \cite{antonini12}, and see a brief summary in \cite{chen19}). In this case,
the SMBH could induce not only a relatively high speed to the BBH but also a
significant gravitational redshift to the GW signal, if the distance between
the BBH and the SMBH is small.  Third, the immediate vicinity of a BBH is not
vacuum either. For example, BBHs may for in a gaseous environment as several
major formation channels would suggest (see a brief review in
\cite{chen20gas}). The interaction with gas could affect the evolution of the
binary too, so that the GW signal may not faithfully reflect the orbital
dynamics in a vacuum. Last but not least, GWs do not propagate freely from the
source to the observer. In between lies structures of a variety of scales and
masses. One well-known effect related to the propagation of GWs in a
non-uniform background is gravitational lensing
\cite{wang96,nakamura98,takahashi03lensing}.  It would distort our perception
of the distance and power of a GW source.

What would happen if one, unaware of the above astrophysical factors, insists
on using the vacuum solution of isolated BBH mergers to model the GW signals?
Could the observer still detect a signal? If a signal is detected, how
accurately could the physical parameters such as mass be retrieved? This
chapter is dedicated to addressing these questions.

\section{Standard siren\index{standard siren}}

In our daily lives, we are accustomed to perceiving the distance of a source
not only by its emitted light but also by its sound.  For example, most of us
have heard the buzz of a bee. We could tell roughly how far it is when we hear
one.  We are able to do this because knowing what a bee sounds like allows us
to calibrate the distance based on the loudness of the buzz. 

More experienced people can even infer the size of the source from the tone of
its sound.  Many of us can distinguish the buzz of a bee from the whining sound
of a mosquito because we know the trick that mosquitoes make higher tones.
Whenever we hear mosquitoes, we know immediately that they are ``dangerously''
close to us. We know it because mosquitoes are not as ``powerful'' as bees, and
hence should be relatively close by once heard.

Analysing GW signals is strikingly similar to perceiving sound. First, lower GW
frequencies normally correlate with more massive objects. Second, once the mass
could be estimated, e.g., from the frequency, we could further infer the
distance according to the loudness, or ``amplitude'' in more physical terms, of
the GW signal.

Among the long list of GW sources, BBHs, binary neutron stars (BNSs), and
double white dwarfs (DWDs) are particularly useful for distance measurement.
This is because their orbital dynamics is simple enough so that we can have a
good theoretical understanding of what their GW radiation is like. In
other words, we know how they ``sound''. Such a binary is often
referred to as a ``standard siren'' \citep{schutz86}.

\begin{figure}
\centering
\includegraphics[width=\textwidth]{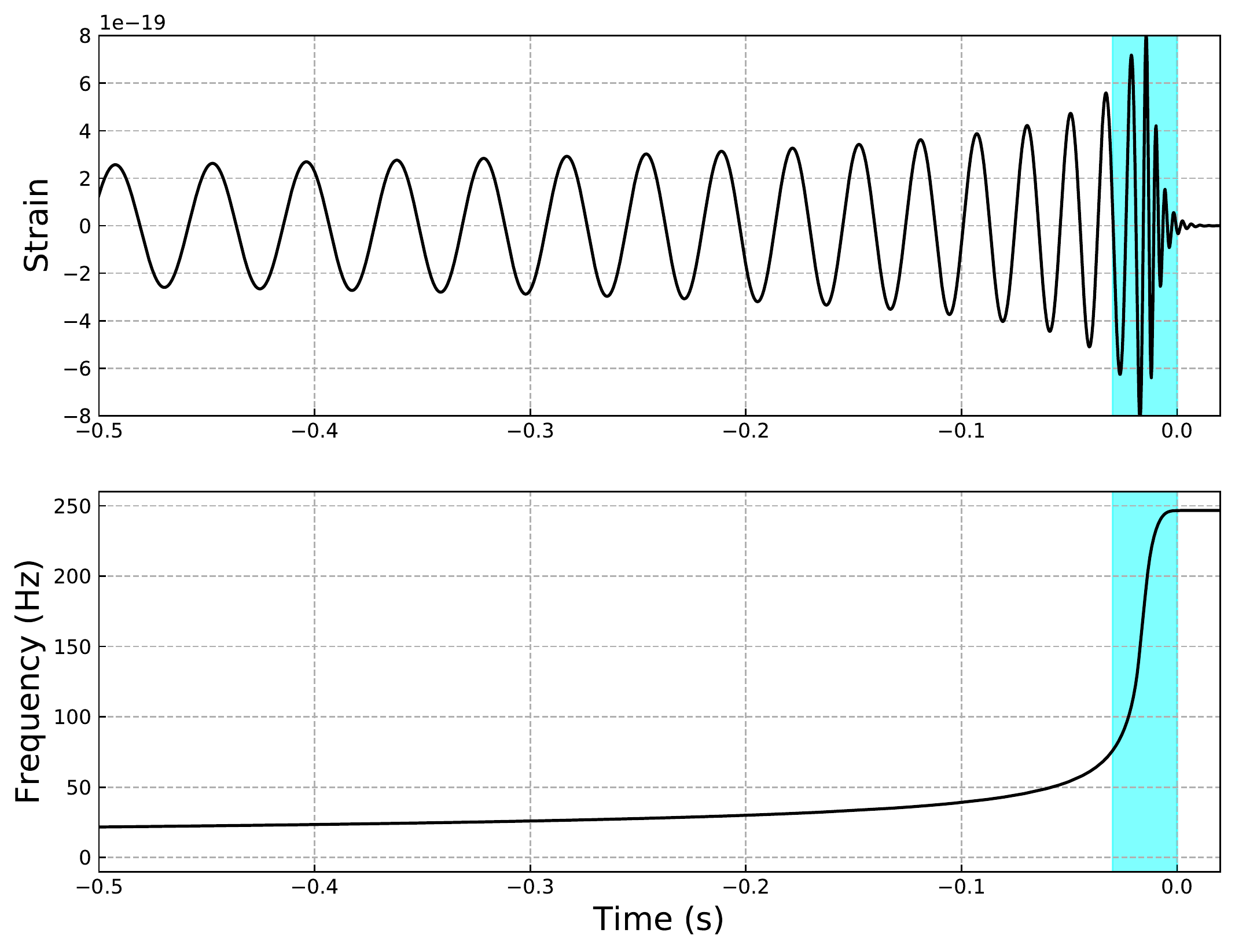}
	\caption{Evolution of the amplitude (upper panel) and frequency (lower
panel) of the GWs produced by a GW150914-like BBH.  In the
calculation, the two BH masses are chosen to be $m_1=40\,M_\odot$ and
$m_2=30\,M_\odot$, and the distance is set to $d=400$ Mpc.  }
\label{fig:signal} 
\end{figure}

To have an idea of what a standard siren is like, we show in
Figure~\ref{fig:signal} the GW signal of a BBH merger similar to GW150914
(adopted from the GW Open Science
Center\footnote{https://www.gw-openscience.org/tutorials/}).  The upper panel
shows the evolution of the amplitude of the space-time distortion,  called
``strain''.  The lower panel shows the corresponding GW frequency.  The merger
of the two BHs happens at the time $t=0$.  We can roughly divide the signal
into three parts. (i) Long before the merger, both the amplitude and the
frequency increase with time. Such a behavior is called ``chirp''\index{chirp}. The physical
reason is that GW radiation causes the orbit to decay, so that the orbital
frequency rises and the acceleration of the mass quadrupole moment intensifies.
During this phase, the orbital semimajor axis is much greater than the
gravitational radii of the BHs.  Therefore, the orbit can be approximated by a
Keplerian motion and the GW radiation can be computed analytically. Since the
shrinkage of the orbit happens on a timescale much longer than the orbital
period, the two BHs spiral inward gradually.  For this reason, this phase is
known as the ``inspiral phase''. (ii) As the separation of the two BHs becomes
comparable to a few gravitational radii, the Keplerian approximation breaks
down and the radial motion of the two BHs becomes more prominent. We can no
longer solve the evolution analytically but have to resort to numerical
relativity. This phase is called the ``merger phase''. It is marked by the cyan
stripes in Figure~\ref{fig:signal}. (iii) Immediately after the coalescence of
the horizons of the two BHs, the remnant is highly perturbed relative to a
stationary Kerr metric. To get rid of the excessive energy, the BH will emit
GWs at a series of characteristic frequencies (not shown in the plot)
determined by the mass and the spin parameters. This process is called
``ringdown''.  Eventually, a single spinning BH is left.

Now that we know what the signal is like after the parameters are specified, 
can we reverse the process and estimate the parameters based on
the signal? In the canonical scenario, i.e., the BBH is isolated in a vacuum
background and not moving relative to us, the answer is yes.

Take the inspiral phase for example (since the chirp signal can be computed
analytically).  We can derive from the signal at least two measurable
quantities, the amplitude $h$ and the frequency $f$. From the evolution of the
frequency, we can further derive the time derivative $\dot{f}$. These three
observables, $h$, $f$, and $\dot{f}$, encode the mass and distance of the
source.

Since both $f$ and $\dot{f}$ are functions of the masses of the two BHs,
$m_1$ and $m_2$, and the semimajor axis of the orbit,  $a$,  
we can combine them to eliminate the dependence on $a$.  We will show that
the result is
a quantity characterizing the mass of the system. First, we follow the Keplerian approximation,
normally acceptable for the inspiral phase,
and derive the GW frequency using twice the orbital frequency,
\begin{equation}
	f= \frac{1}{\pi}\left[\frac{G(m_1+m_2)}{a^3}\right]^{1/2},\label{eq:f}
\end{equation}
where $G$ is the gravitational constant. Here we have assumed a circular orbit for
simplicity. Elliptical orbits will lead to additional harmonics of different frequencies.
Because of the GW radiation, the semi-major axis $a$ decays as
\begin{equation}
        \dot{a}_{\rm gw}:=-\frac{64}{5}\frac{G^3m_1m_2(m_1+m_2)}{c^5a^3},\label{eq:adotgw}
\end{equation}
where $c$ is the speed of light (as is derived in \cite{peters64}). Combining the previous two equations,  we can derive
\begin{equation}
	\dot{f}=\dot{f}_{\rm gw}=\frac{96G^{7/2}m_1m_2(m_1+m_2)^{3/2}}{5\pi c^5a^{11/2}},\label{eq:fdotgw}
\end{equation}
where $\dot{f}_{\rm gw}$ denotes the contribution from GW radiation only.
Now we can eliminate the $a$ in Equation~(\ref{eq:fdotgw}) using Equation~(\ref{eq:f}).
The result is
\begin{equation}
	\frac{c^3}{G}\,\left(\frac{5f^{-11/3}\dot{f}}{96\pi^{8/3}}\right)^{3/5}=\frac{(m_1m_2)^{3/5}}{(m_1+m_2)^{1/5}}.\label{eq:chirpmass}
\end{equation}
It is now clear that from the two observables $f$ and $\dot{f}$, one can derive a quantity of the dimension of mass,
\begin{equation}
	{\cal M}=\frac{(m_1m_2)^{3/5}}{(m_1+m_2)^{1/5}}.\label{eq:M}
\end{equation}
This mass is knowns at the ``chirp mass''\index{chirp mass} 
because, as Equation~(\ref{eq:chirpmass}) indicates,
it determines how $f$ increases with time.

Besides the chirp mass, can we derive $m_1$ and $m_2$ separately?  It is
difficult if the orbit is nearly Keplerian. In this case, we have seen that
$\dot{f}$ is determined only by $\cal{M}$. The mass ratio of the two BHs,
$q:=m_2/m_1$ (assuming $m_2\le m_1$), plays no role as long as $\cal{M}$ is
fixed.  However, as the binary shrinks to a size of about dozens of the
gravitational radii, GR effect becomes significant and the orbit starts to
deviate from Keplerian motion. In fact, the deviation is
$q$-dependent. Therefore, we can use it to disentangle the masses of
the two BHs.

\begin{figure}
\centering
\includegraphics[width=0.9\textwidth]{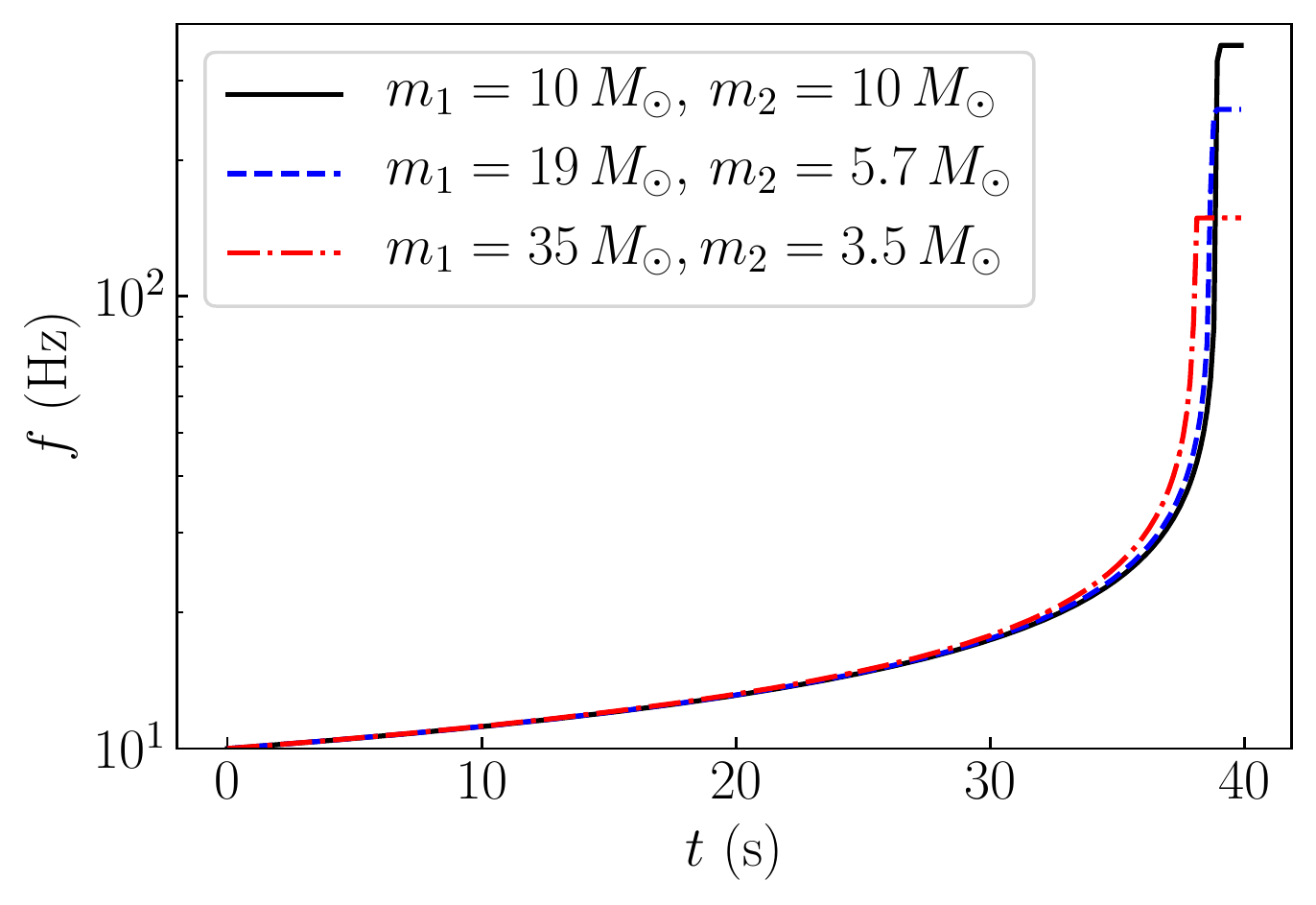}
\caption{Chirp signals of three BBHs with the same chirp mass, ${\cal M}\simeq8.7$,
but different mass ratios, $q=1$, $1/3$, and $1/10$. The effect of different mass ratio
starts to appear when $f$ exceeds about $20$ Hz.}
\label{fig:chirp}
\end{figure}

To illustrate this idea, we show in Figure~\ref{fig:chirp} the chirp signals of
three BBHs with the same chirp mass but different mass ratios. To include the
GR effects, the waveforms are computed to the 3.5 post-Newtonian order
\cite{sathyaprakash09}. Since the chirp masses are the same, the three signals
are almost identical when $f\lesssim15$ Hz. At the later evolutionary stage
when $f\gtrsim20$ Hz, we start to see a divergence.  The most
prominent difference is the final frequency. We can see that higher frequency
is correlated with larger mass ratio ($q$ approaches $1$). This is because as
$q$ increases, the mass of the primary BH ($m_1$) decreases, so that the inner
most circular orbit (ISCO\index{ISCO}) is smaller, and the corresponding frequency is
higher.

This general result suggests that one has to detect the final merger to be able
to measure the mass ratio of a BBH. This is the reason that LIGO and Virgo, by
detecting GWs in the $10-10^2$ band, are capable of estimating the masses of
individual BHs in binaries. The Laser Interferometer Space Antenna (LISA, see
\cite{amaro17}), on the other hand, can detect only the early inspiral phase of
(stellar-mass) BBHs because LISA is tuned to be sensitive to milli-Hz (mHz)
GWs. As a result, it is generally more difficult for LISA to measure the
individual masses of (stellar-mass) BHs. It is worth noting that LISA is
capable of measuring the masses of intermediate-massive and supermassive BHs between
$10^3$ and $10^{10}\,M_\odot$, because their final mergers produce mHz GWs.

How about measuring the distance? So far, we have not yet used the third observable $h$. 
In principle, it is inversely proportional to the distance because larger distance should
lead to smaller GW amplitude. In fact, in the Keplerian approximation the distance 
$d$ depends on $h$ as
\begin{equation}
d=\frac{4G}{c^2}\frac{{\cal M}}{h}\left(\frac{G}{c^3}\pi\,f\,{\cal M}\right)^{2/3}.\label{eq:d}
\end{equation}
Here we have omitted the uncertainty induced by the inclination of the binary orbit because
in principle it can be eliminated by measuring the GW polarizations.
The important point is that all the quantities on the right-hand-side of the
last equation can be derived from the three observables $f$, $\dot{f}$, and
$h$. This completes the theory of measuring the mass and distance of a BBH
using the chirp signal. 

\section{Mass-redshift degeneracy\index{degeneracy}}

\begin{figure}
\centering
\includegraphics[width=0.9\textwidth]{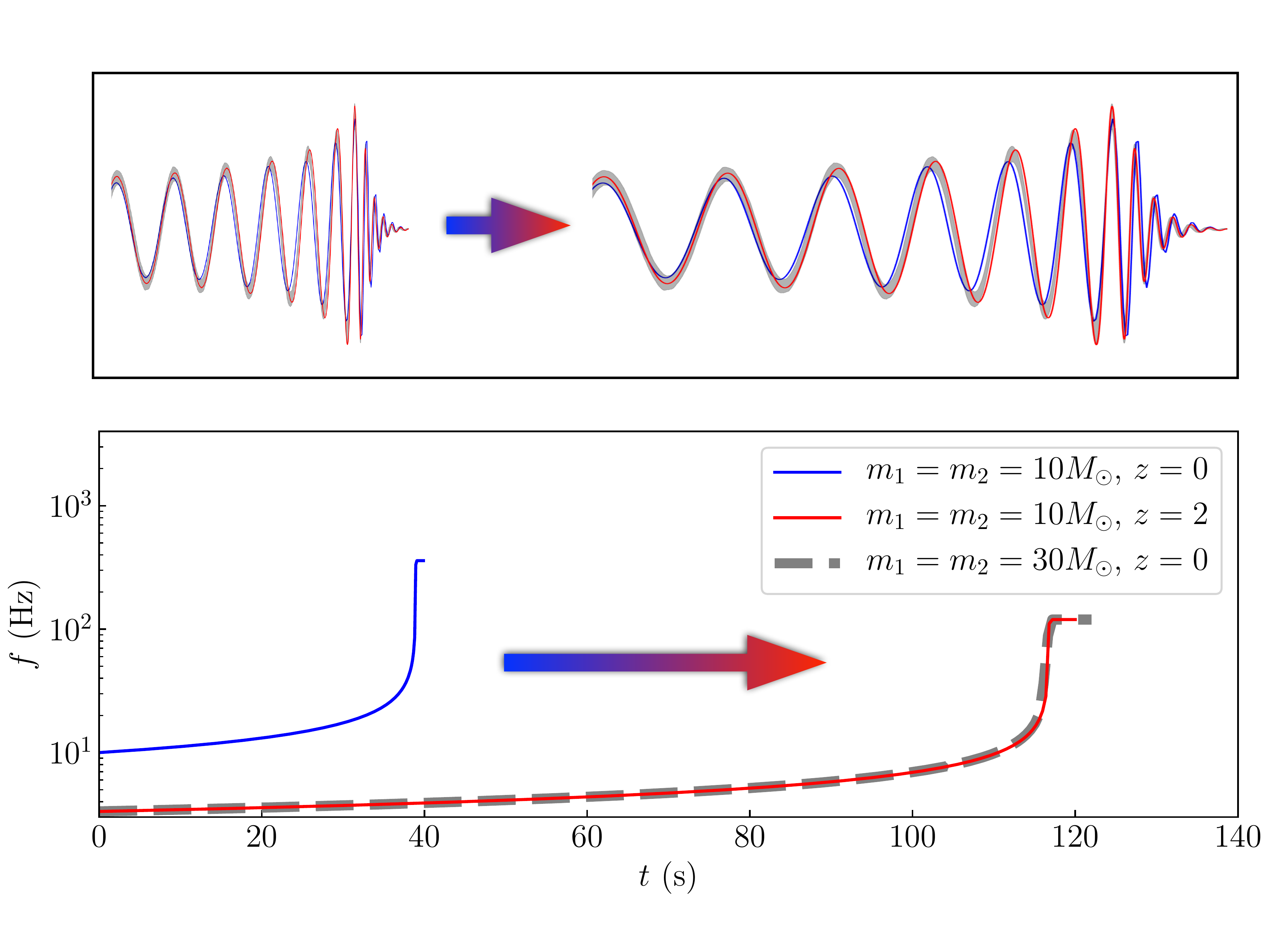}
\caption{Effects of redshift on GW signals. Upper panel: Redshift stretches a waveform. 
Lower panel: An illustration of the problem of mass-redshift degeneracy. The intrinsic chirp signal of a $10-10\,M_\odot$ binary (blue solid curve) gets 
redshifted if the binary is placed at a redshift of $z=2$ (red solid curve). The redshifted signal
is identical to that of a more massive, $30-30\,M_\odot$ binary residing at $z=0$ (grey dashed curve).
}
\label{fig:redshift}
\end{figure}

In the last section, we have assumed a flat, Minkowski space. Because of this
assumption, the signal detected is identical to the signal emitted from the
source. However, the expansion of the universe will complicate this
relationship. The cosmological expansion effectively stretches the signal.
Consequently, the detected GWs will appear redshifted.
Figure~\ref{fig:redshift} illustrates this effect. From such a distorted
signal, can we still retrieve the correct mass and distance of the source? 

As a result of the cosmological redshift, the apparent frequency $f_o$ 
will be lower than the 
intrinsic frequency $f$ measured in the rest frame of the source. The difference 
can be calculated with
\begin{equation}
	f_o=f(1+z_{\rm cos})^{-1},\label{eq:f_o}
\end{equation}
where $z_{\rm cos}$ denotes the cosmological redshift.
Moreover, to the observer the chirp rate appears to be 
\begin{equation}
	\dot{f}_o=\dot{f}_{\rm gw}(1+z_{\rm cos})^{-2},
\end{equation}
where the additional factor of $(1+z_{\rm cos})^{-1}$ is due to time dilation\index{time dilation}.
The students can verify the above equation by noticing that
$\dot{f}_o=df_o/dt_o$ and $dt/dt_o=(1+z_{\rm cos})^{-1}$.
If we wish to derive a chirp mass from the observed, redshifted signal, we can only
get
\begin{equation}
{\cal
	M}_o=\frac{c^3}{G}\,\left(\frac{5f_o^{-11/3}\dot{f_o}}{96\pi^{8/3}}\right)^{3/5}={\cal M}(1+z_{\rm cos}).\label{eq:M_o}
\end{equation}
This new mass is greater than the intrinsic chirp mass by a redshift factor
$1+z_{\rm cos}$. For this reason,  it is called the ``redshifted chirp mass''.

Now we have a problem. The mass appears greater than the intrinsic one.
More seriously, without knowing the cosmological redshift of the source, we
would not know the real mass. This famous problem in GW astronomy
is called ``the mass-redshift degeneracy''.  Given such a problem,  should we
trust the BH masses derived from GWs, especially when the detected BHs seem a
bit overweight?

There are two possibilities of breaking this degeneracy.
One possibility is to first measure the distance
using GWs and then infer the redshift based on a chosen cosmological model. But what kind of
distance is measurable from GWs in an expanding universe?  Let us take the $\Lambda$CDM
cosmology for example ($h=0.7$, $\Omega_M=0.29$, and $\Omega_\Lambda=0.71$). 
In this cosmology, the GW amplitude an observer detects
is related to the transverse comoving distance $d_C$ as
\begin{equation}
h_o=\frac{4G}{c^2}\frac{{\cal M}}{d_C}\left(\frac{G}{c^3}\pi\,f\,{\cal M}\right)^{2/3},\label{eq:h_o}
\end{equation}
where $f$ and $\cal{M}$ refer to the intrinsic, 
non-redshifted quantities. Using this apparent GW amplitude, we can only infer
a distance of
\begin{equation}
d_o=\frac{4G}{c^2}\frac{{\cal M}_o}{h_o}\left(\frac{G}{c^3}\pi\,f_o\,{\cal M}_o\right)^{2/3}.\label{eq:d_o}
\end{equation}
Substituting the observed quantities on the right-hand-side (those with the subscript $o$) 
using Equations~(\ref{eq:f_o}), (\ref{eq:M_o}), and (\ref{eq:h_o}), one will find that
\begin{equation}
d_o=d_C(1+z_{\rm cos})=d_L,
\end{equation}
where $d_L$ is the luminosity distance\index{luminosity distance}. 
Therefore, what we derive from a chirp signal 
is in fact the luminosity distance. From the luminosity distance, we
can calculate the corresponding redshift based on the chosen cosmological
model. Finally, with this redshift, we can break the mass-redshift degeneracy.

\begin{figure}
\centering
\includegraphics[width=\textwidth]{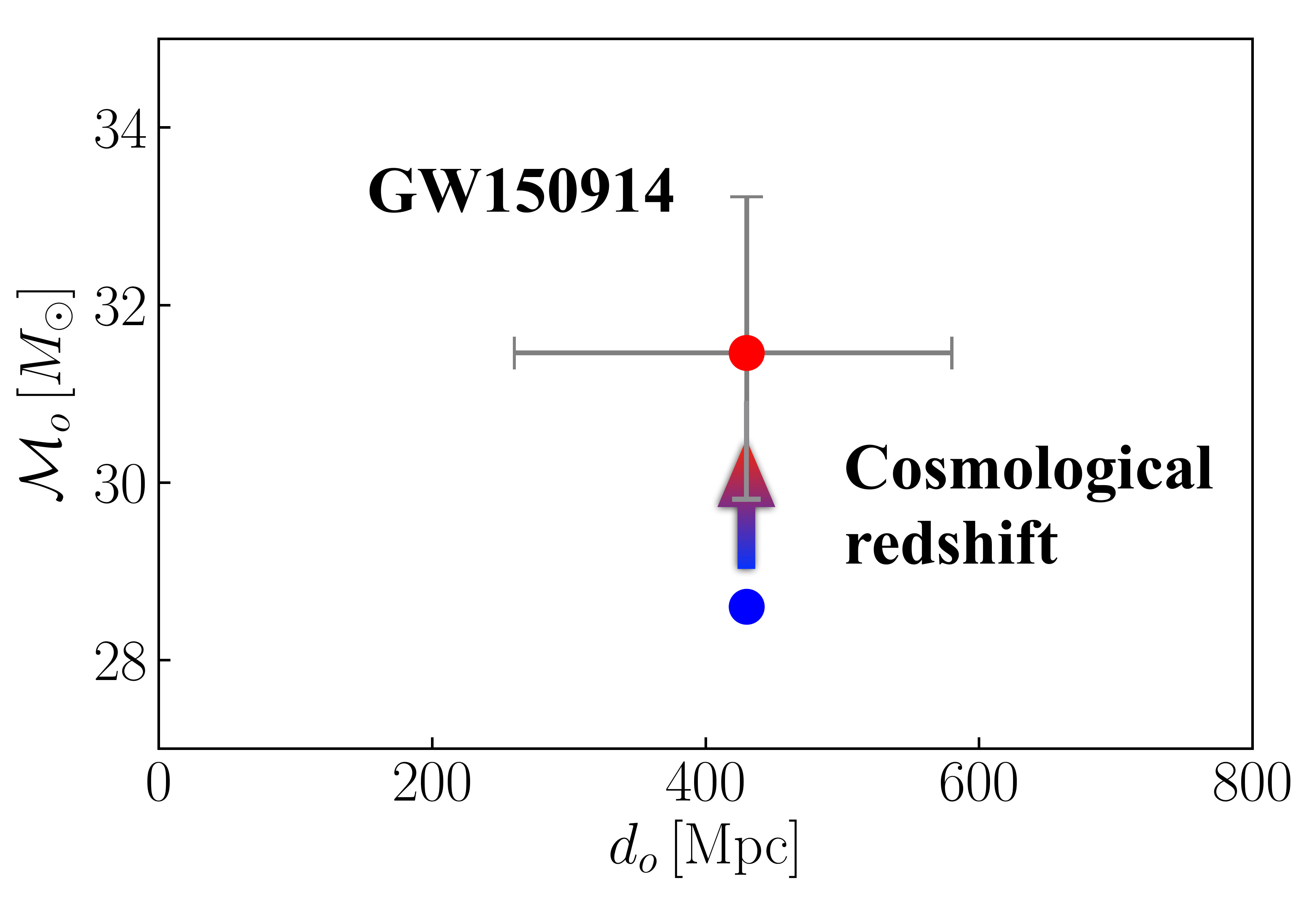}
\caption{The effect of cosmological redshift on the apparent mass and distance of GW150914.
The blue dot marks the location of the real chirp mass and (luminosity) distance of the source.
The red dot shows the redshifted chirp mass and apparent distance, measureable from
GW signal.}
\label{fig:cos}
\end{figure}

To illustrate this idea, we show in Figure~\ref{fig:cos} the effect of
cosmological expansion on the apparent chirp mass and distance of the first GW
event GW150914. In the $\Lambda$CDM cosmology we have chosen, the apparent
distance $d_o$ is identical to the luminosity distance, which corresponds to a
redshift of $z_{\rm cos}\simeq0.09$.  Therefore, we can infer that the
intrinsic chirp mass ${\cal M}$ is smaller than the apparent one, ${\cal M}_o$,
by only a factor of about $1.1$. For this reason, it is believed that the BHs
in GW150914 are intrinsically massive.

Another possibility is that an electromagnetic counterpart associated with a GW
event is also detected.  Then the redshift can be measured independently, e.g.,
from the spectrum. Detecting the electromagnetic counterpart will greatly
enhance the scientific payback of GW observations. With the redshift $z_{\rm
cos}$ derived from the counterpart and the luminosity distance $d_L$ from the
GW signal, one can measure the Hubble constant and, in turn, constrain
cosmological models.  Unfortunately, detecting electromagnetic counterparts is
challenging with the current telescopes. Moreover, some GW events, such as BBHs
merging in a vacuum background, are not expected to produce strong
electromagnetic radiation.

\section{Effects of astrophysical environments}

The standard model of a BBH merger does not take into account the motion of the
binary relative to the observer. Neither does it include other
astrophysical objects, or matter around the BHs or along the path of the
propagation of GWs. Now we will add these new ingredients into our model and
study the potential impact on our measurement of
the physical parameters of the source.

\subsection{Strong gravitational lensing\index{lensing}}

Strong gravitational lensing by foreground galaxies or galaxy clusters has been
observed for transient light sources, such as supernovae and gamma-ray bursts.
It is reasonable to speculate that some LIGO/Virgo events may also
be lensed. 

The effect of strong lensing is to magnify the amplitude of GWs. The
significance can be characterized by the magnification\index{magnification}
factor ${\cal A}$, such that
\begin{equation}
h_o={\cal A}\left(\frac{4G}{c^2}\frac{{\cal M}}{d_C}\right)
\left(\frac{G}{c^3}\pi\,f\,{\cal M}\right)^{2/3}.
\end{equation}
According to Equation~(\ref{eq:d_o}), the distance of a lensed GW sources will appear to be
\begin{equation}
d_o=d_L/\cal{A},
\end{equation}
which is smaller than the real luminosity distance.
The apparent mass, which one would derive from the GW signal, remains 
\begin{equation}
	{\cal M}_o={\cal M}(1+z_{\rm cos}).
\end{equation}

\begin{figure}
\centering
\includegraphics[width=\textwidth]{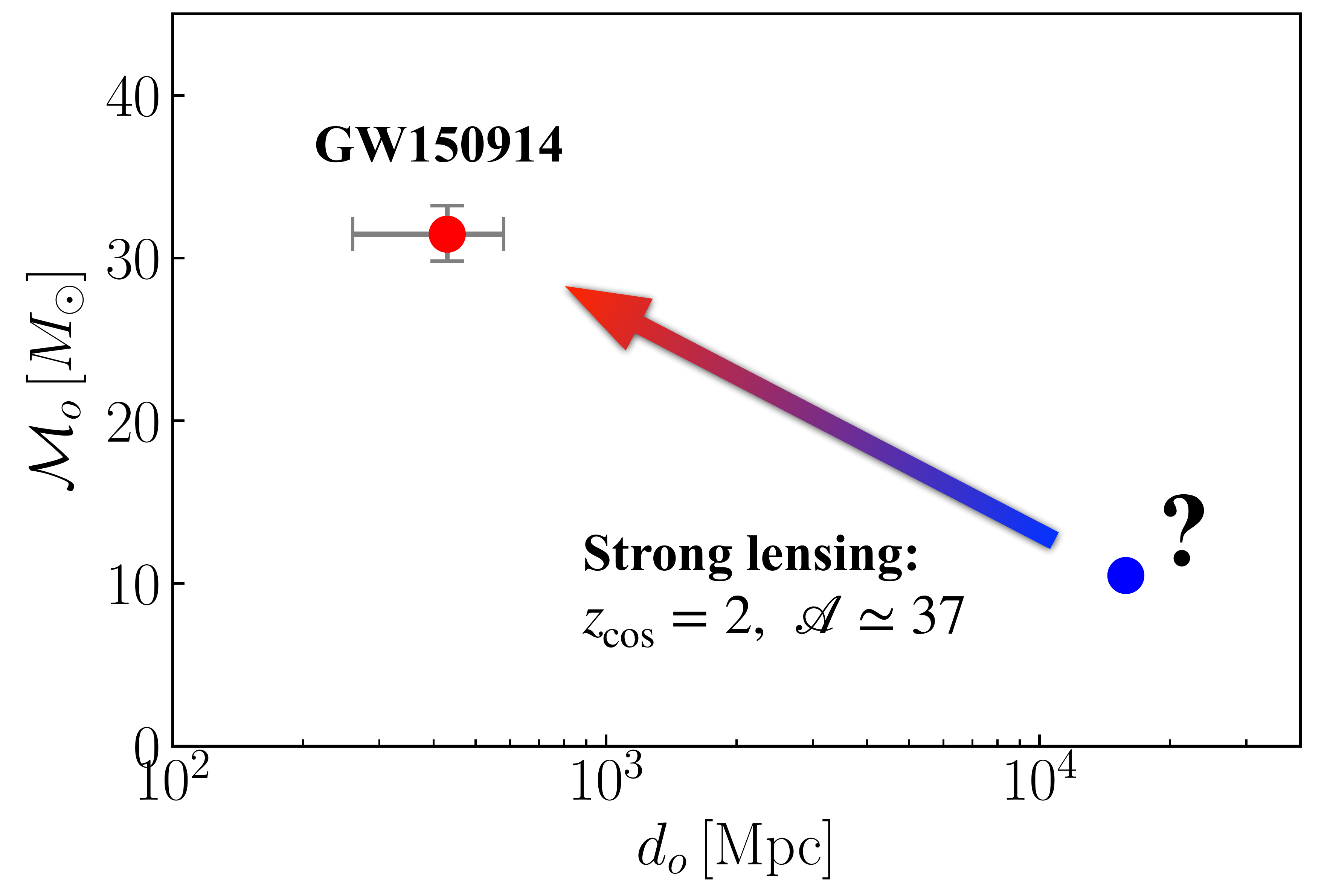}
\caption{The lensing model of GW150914. 
For illustrative purposes, 
the source (blue dot) is placed at a cosmological redshift of $z_{\rm cos}=2$, 
corresponding to a luminosity distance of $16$ Gpc in the $\Lambda$CDM cosmology.
After lensing, 
the mass and distance appear to coincide with those of GW150914 (red dot). The corresponding
magnification factor
is ${\cal A}\simeq37$.}
\label{fig:lense}
\end{figure}

Figure~\ref{fig:lense} illustrates this effect using GW150914 as an example.
It shows that a low-mass BBH residing at a high redshift, if getting
gravitationally lensed, would appear as a high-mass binary residing at a
relatively low redshift (as has been pointed out in
\cite{broadhurst18,smith18}).  Now we cannot use the apparent distance $d_o$ to
break the mass-redshift degeneracy, because it does not reflect the real
luminosity distance of the source.

Could strong lensing explain all the massive BBHs detected by LIGO and Virgo?
It is difficult because from the statistical point of view, lensing events
should be uncommon. Moreover, as is indicated by the direction of the arrow in
Figure~\ref{fig:lense}, the lensing effect tends to produce an anti-correlation
between the observed quantities $d_o$ and ${\cal M}_o$ \cite{broadhurst18}. The
current observations do not favor such a relation \cite{abbott19GWTC}.
However, as the number of LIGO/Virgo BBHs increase and the detection horizon
expands, the possibility of catching a lensed event increases as well.
Therefore, watch out for massive BBHs at relatively low redshift!

\subsection{The effect of motion}

Almost all astrophysical objects are moving relative to us. The general effect
of motion is to induce
a Doppler shift\index{Doppler shift} to the frequency of a wave signal. This shift can be characterized by
the relativistic Doppler factor\index{Doppler factor},
\begin{equation}
	1+z_{\rm dop}=\gamma(1+\vec{v}\cdot\vec{n}/c),
\end{equation}
where $\vec{v}$ is the velocity of the source, $\vec{n}$ is a unit vector
coinciding with the line-of-sight, and $\gamma=1/\sqrt{1-(v/c)^2}$ is the
Lorentz factor. 

Using this factor and taking the cosmological redshift into account, the apparent frequency can be written as
\begin{equation}
	f_o=f(1+z_{\rm cos})^{-1}(1+z_{\rm dop})^{-1},\label{eq:fodop}
\end{equation} 
and the chirp rate changes to
\begin{equation}
	\dot{f}_o=\dot{f}_{\rm gw}(1+z_{\rm cos})^{-2}(1+z_{\rm dop})^{-2}.\label{eq:fodotdop}
\end{equation}
Applying Equation~(\ref{eq:M_o}) again, we find that the 
apparent chirp mass is
\begin{equation}
	{\cal M}_{o}={\cal M}(1+z_{\rm cos})(1+z_{\rm dop}).
\end{equation}
As for the distance, we notice that motion does not significantly affect the amplitude of
GWs \cite{torres-orjuela19}. Therefore, we can follow Equation~(\ref{eq:h_o})
to calculate $h_o$, and further derive that 
\begin{equation}
        d_{o}=d(1+z_{\rm cos})(1+z_{\rm dop}).
\end{equation}

The last two equations suggest that Doppler shift can also affect the measurement
of the mass and distance of BBH. It introduces a systematic error if not
appropriately accounted for in GW data analysis.  In particular, when the
relative velocity and the line-of-sight are aligned, i.e.,
$\vec{v}\cdot\vec{n}>0$, the Doppler effect would make the mass and distance
appear even greater than their intrinsic values. 

We note one difference between the Doppler effect and the cosmological
redshift.  While cosmological redshift always makes the mass and distance
appear bigger, Doppler effect could make them appear smaller as well.  This
happens when the source is moving towards the observer, i.e., when
$\vec{v}\cdot\vec{n}<0$. In this case, the Doppler factor, as well as the
effective redshift factor $1+z_{\rm dop}$, becomes smaller than $1$.

How large could the factor $1+z_{\rm dop}$ be?  In the conventional scenario,
BBHs form either in isolated binaries in galaxy bulges and disks, or in star
clusters \cite{ligo16astro}. In our Milky Way, isolated binaries and star
clusters are moving at a typical velocity of $200\,{\rm km\,s^{-1}}$ relative
to the earth. Therefore, their Doppler effect can be neglected.  
For the BBHs in external galaxies, their velocities relative to the
Milky Way could be much greater. For example, in the most massive galaxy
clusters, the velocities of the galaxy members can reach thousands of ${\rm
km\,s^{-1}}$. Even this velocity is much smaller than the speed of light.  For
these reasons, Doppler effect is neglected in the standard procedure of GW data
analysis.

However, the conventional picture of BBH formation is incomplete. Recent
studies suggest that in galaxy centers, especially where there are SMBHs, the
merger rate of BBHs can be enhanced.  The reasons are multifold \cite{chen19}.
(i) Close to a SMBH, the escape velocity is large.  This is a place where newly
formed compact objects are likely to stay. (ii) SMBHs are normally surrounded
by a cluster of stars, known as the ``nuclear star cluster''. Due to a net loss
of energy during the gravitational interaction with other stars, massive
objects such as BHs and neutron stars will gradually segregate towards
the center of a nuclear star cluster. This ``mass segregation\index{mass segregation} 
effect'' increases even higher the
concentration of compact objects near SMBHs. Such a condition is favorable
to the formation of BBHs. (iii) The tidal force exerted by SMBHs could excite
the eccentricities of the nearby BBHs through the so-called ``Lidov-Kozai''
mechanism. A higher eccentricity normally results in a stronger GW radiation,
and hence leads to a faster merger.  (iv) If a SMBH is surrounded by
gas, as would be the case in an active galactic nucleus (AGN) due to the
presence of a gaseous accretion disk, interaction with the gas could also
dissipate the internal energy of a BBH and lead to its faster merger. 

Therefore, it is possible that a BBH may merge in the vicinity of a SMBH.  If
such a merger happens, the binary is likely to obtain a large velocity due to
the orbital motion around the SMBH.  For a BBH detected by LIGO/Virgo, the velocity is
almost constant  because the event lasts normally less than a second but the
orbital period around the SMBH is orders of magnitude longer. The tidal
force of the SMBH does not significantly affect the shrinkage of the binary
because GW radiation predominates at this stage.  Under these conditions, 
the theory developed in this section applies. 

It is clear that the effect is more significant if the merger happens closer to
the SMBH. We can use the velocities at the last stable orbits to estimate the
order of magnitude of the maximum effect.  If the binary is on a circular orbit around
the SMBH, as we would expect for the BBHs in AGN accretion disks, the largest velocity
appears at the ISCO\index{ISCO}.  For non-spinning Schwarzschild BHs, the orbit is at a
radius of $R=6GM/c^2$, where $M$ is the mass of the SMBH. In this case, the
velocity is approximately $c/\sqrt{6}\simeq0.408c$. The corresponding Doppler
factor is $1+z_{\rm dop}\simeq1.54$. On the other hand, if the orbit of the BBH
is nearly parabolic, a more likely configuration if the binary forms far away
and later gets captured by the SMBH, the last orbit is the innermost
bound orbit (IBO\index{IBO}). In this case, the pericenter is at $4GM/c^2$. The
corresponding velocity is $c/\sqrt{2}=0.707c$ and the Doppler factor becomes
$1+z_{\rm dop}\simeq2.41$.  Of course, even higher values are possible
if the
distance is smaller, but such events are rarer.  The values derived above
suggest that the Doppler effect could significantly affect our measurement of
the mass and distance of a GW source, if the source is close to a SMBH. The event rate,
therefore, deserves careful calculation.

\subsection{Deep gravitational potential}

Close to a SMBH, gravitational redshift is also significant. This
additional redshift would further distort a GW signal and make the measurement
of the physical parameters even more biased.

Suppose the gravitational redshift is $z_{\rm gra}$, the observed GW frequency 
would be lowered relative to
the emitted one by a factor of $1+z_{\rm gra}$. Similar to the cosmological
redshift and the Doppler effect, the chirp rate would be reduced by a factor of
$(1+z_{\rm gra})^2$ due to time dilation. Using these relations, it is straightforward
to show that the chirp mass appears to be
\begin{equation}
{\cal M}_o={\cal M}(1+z_{\rm cos})(1+z_{\rm dop})(1+z_{\rm gra})
\end{equation} 
to an observer. To derive the apparent distance, we notice that
the GW amplitude is not significantly affected by the gravitational redshift
as long as the merger does not happen extremely close to the event horizon (see \cite{chen19} for
a more quantitative description). As a result, 
Equation~(\ref{eq:h_o}) remains a good approximation, from which we can derive that
\begin{equation}
	d_o=d_C(1+z_{\rm cos})(1+z_{\rm dop})(1+z_{\rm gra}).
\end{equation}
We can see that the apparent mass and distance of a BBH are determined by three
types of redshift. Nevertheless, LIGO/Virgo only considered the cosmological redshift
in their analysis. Such a treatment is not suitable for the BBHs forming in
the vicinity of SMBHs. 

We can again evaluate how extreme the value of $(1+z_{\rm dop})(1+z_{\rm gra})$
could be by investigating the last stable orbits.  Assuming a non-spinning
SMBH, we can calculate the gravitational redshift\index{gravitational redshift} by
\begin{equation}
1+z_{\rm gra}=(1-R_S/R)^{-1/2},
\end{equation}
where $R_S=2GM/c^2$ is the Schwarzschild radius. Using this equation, we find that
for ISCO, the value of $(1+z_{\rm dop})(1+z_{\rm
gra})$ is approximately $1.89$, while for IBO, it is
about $3.41$. 

\begin{figure}
\centering
\includegraphics[width=0.9\textwidth]{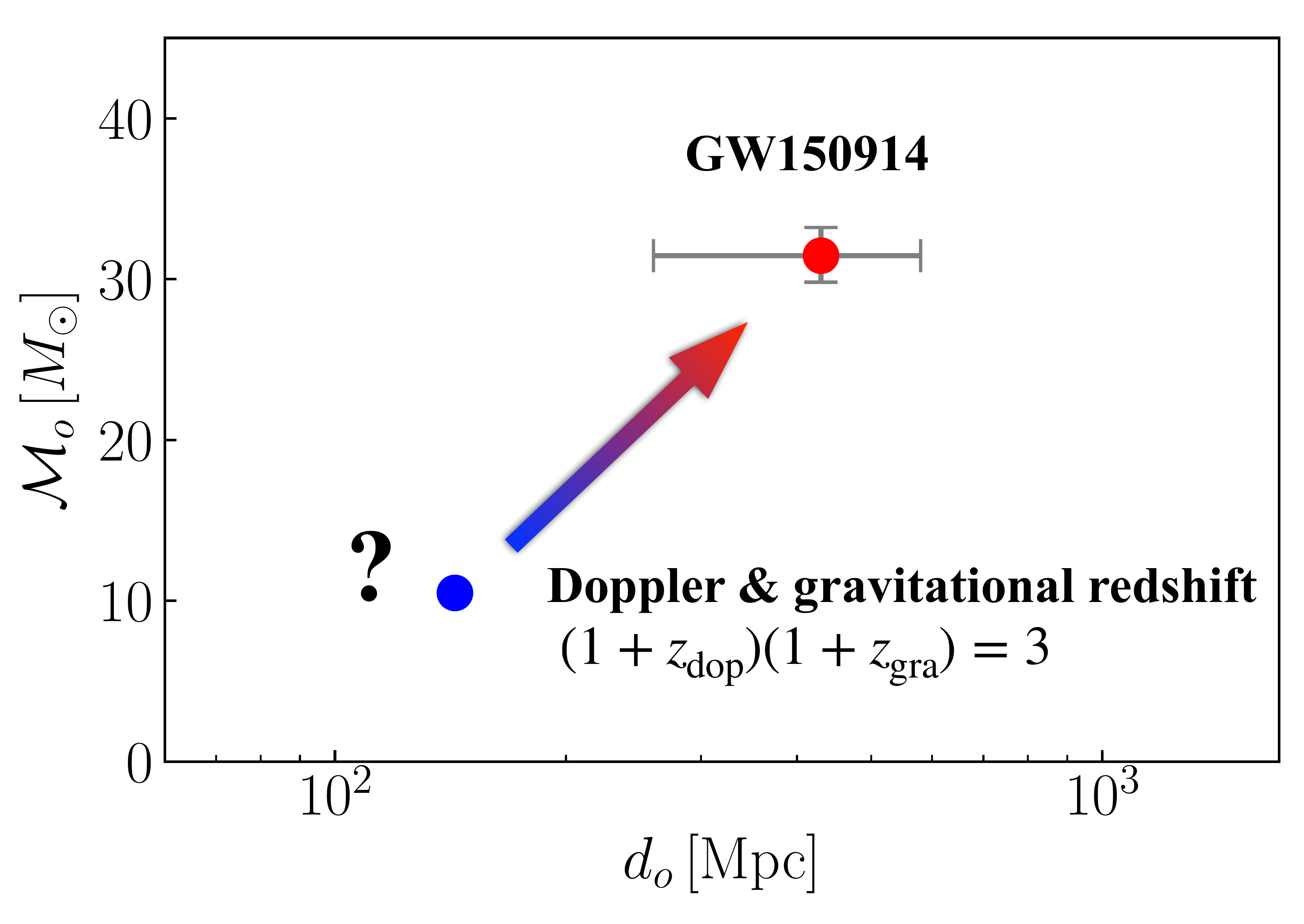}
\caption{The Doppler-plus-gravitational-redshift model for GW150914. The location of the
intrinsic chirp mass and luminosity distance is marked by the blue dot. The location of the
apparent mass and distance is shown by the red dot. In the plot, a total redshift factor of 
$(1+z_{\rm dop})(1+z_{\rm gra})=3$ is assumed, which corresponds to a place close to the
	pericenter of the IBO around a Schwarzschild SMBH.}
\label{fig:dopgra}
\end{figure}

Interestingly, the values of these redshift factors coincide with the
difference between the BH masses in GW150914 ($\sim30\,M_\odot$) and those in
X-ray binaries ($\sim10\,M_\odot$). Could the Doppler$+$gravitational redshift
explain the high-mass BHs detected by LIGO and Virgo (see
Figure~\ref{fig:dopgra})?  The key to answer this question is, again, the event
rate.  More works along this line of research are deserved.

\subsection{Peculiar acceleration\index{peculiar acceleration}}

For a binary orbiting a SMBH or any other tertiary object, the velocity of the
center-of-mass of the binary in principle is not constant. This acceleration
may not be a problem for LIGO/Virgo binaries because the event lasts only a
fraction of a second, too short compared to the orbital period of the tertiary
to cause any noticeable changes of the velocity. However, for LISA BBHs, the
acceleration is no longer negligible. This is because LISA sources could dwell in
the band for years. During this period, the velocity of the binaries may change
significantly. This in turn changes the Doppler redshift \cite{meiron16,inayoshi17}. 

Let us study the effect of this peculiar acceleration 
(different from the acceleration of the cosmological expansion) on our measurement of 
the mass and distance of binary compact objects.
Starting from Equation~(\ref{eq:fodop}) and neglecting the cosmological redshift
for the time being, we can rewrite the relationship between the apparent frequency and
the center-of-mass velocity of the binary as 
\begin{equation}
	f_o=f\frac{\sqrt{1-\beta^2}}{1+\vec{v}\cdot\vec{n}/c},
\end{equation}
where $\beta=v/c$ and $f$ remains twice the orbital frequency. 
Differentiate it relative to the observer's time, $t_o$, we get
\begin{align}
	\dot{f}_o&=\frac{df_o}{dt_o}=\frac{dt}{dt_o}\frac{df_o}{dt}
	=\frac{1}{1+z_{\rm dop}}\left[\frac{\dot{f}_{\rm gw}}{1+z_{\rm dop}}
	+f\frac{d}{dt}\left(\frac{\sqrt{1-\beta^2}}{1+\vec{v}\cdot\vec{n}/c}\right)\right]\\
	&=\frac{\dot{f}_{\rm gw}}{(1+z_{\rm dop})^2}+f_o\frac{d}{dt}\left(\frac{\sqrt{1-\beta^2}}{1+\vec{v}\cdot\vec{n}/c}\right).\label{eq:fodotacc}
\end{align}
If we compare the last equation with Equation~(\ref{eq:fodotdop}) and notice
that $z_{\rm cos}=0$ for now, we will see that the second term on the
right-hand-side of the last equation is caused by the variation of $\vec{v}$.
This term could be either positive or negative, depending on the orbital
dynamics. Therefore, the apparent chirp mass, which should be calculated from
$f_o$ and $\dot{f}_o$, could also be bigger or smaller than the real value, and
so does the apparent distance. To further quantify this effect, we define
$\Gamma_{\rm acc}$ to be the ratio between the second and the first term in
Equation~(\ref{eq:fodotacc}). This treatment allows us to write ${\cal M}_o$
and $d_o$ in the following compact forms,
\begin{equation}
	{\cal M}_o={\cal M}(1+z_{\rm dop})(1+\Gamma_{\rm acc})^{3/5},\label{eq:Moacc}
\end{equation}
and
\begin{equation}
	d_o=d(1+z_{\rm dop})(1+\Gamma_{\rm acc}).
\end{equation}
In particular, Equation~(\ref{eq:Moacc}) indicates that mass is degenerate with
not only redshift but also acceleration. This is a new type of degeneracy in GW
astronomy. Notice that $\Gamma_{\rm acc}$ could be negative so that
$1+\Gamma_{\rm acc}$ may be smaller than $0$. In this case, we see an inverse
chirp, i.e., the GW frequency decreases with time. Such a signal immediately
indicates a non-standard condition for the binary. These events could be
singled out in data analysis without causing further confusion. Therefore, 
in the following we focus on the case in which $1+\Gamma_{\rm acc}>0$.

For those binaries in triple systems, when will $\Gamma_{\rm acc}$ significantly
exceed unity?  Since $\dot{f}_{\rm gw}\propto f^{11/3}$, the value of
$\Gamma_{\rm acc}$ is higher for lower $f$. This is another reason that we focus on LISA
binaries in this section.  To derive a value for $\Gamma_{\rm acc}$, we
must specify the parameters of the tertiary.  For simplicity, we assume
circular orbits for the tertiary. In this case, the second term on the
right-hand-side of Equation~(\ref{eq:fodotacc}) is of the order of
$Gm_3f_o/(r^2c)$, where $m_3$ is the mass of the tertiary and $r$ is the
distance between the tertiary and the center-of-mass of the binary.  Here we have
omitted the term $\sqrt{1-\beta^2}$ because we only consider the case in which
the center-of-mass velocity is much smaller than the speed of light. 
This consideration is acceptable for LISA binaries because they cannot
reside at several Schwarzschild radii of a SMBH. Otherwise, the binaries would be tidally disrupted
given their large semimajor axes. 
Using this
approximation for the acceleration, we find that
\begin{equation}
	\Gamma_{\rm acc}\simeq4.4\left(\frac{f}{2\,{\rm mHz}}\right)^{-8/3}
	\left(\frac{{\cal M}}{0.3\,M_\odot}\right)^{-5/3}
	\left(\frac{m_3}{1\,M_\odot}\right)
	\left(\frac{r}{30\,{\rm AU}}\right)^{-2}.
\end{equation}

This result suggests that the acceleration affects more seriously the low-mass
binaries (small ${\cal M}$) in the LISA band.  Let us take DWDs
for example since they are the most numerous binaries in band \cite{amaro17}.
If we assume ${\cal M}=0.3\,M_\odot$ and set $f=2$ mHz, the most sensitive band
of LISA, we find that $\Gamma_{\rm acc}=4.4$ for a tertiary of $m_3=1\,M_\odot$
and  $r=20$ AU. Corresponding, the chirp mass appears bigger, due to the
term $(1+\Gamma_{\rm acc})^{3/5}$, by a factor of $2.8$. The distance, on the
other hand, appears bigger by a factor of about $5.4$. In this case, a DWD inside
the Milky Way would appear in a nearby galaxy! If, on the other hand, we reverse
the orbital velocity of the tertiary and increase the chirp mass to $0.8\,M_\odot$,
we will get $\Gamma_{\rm acc}=-0.86$. In this case, the DWD will appear lighter by a
factor of $3.2$ and closer by a factor of $7.0$! Similar cases could potentially
impede our understanding of the formation and distribution of DWDs in the Milky Way.

If the tertiary is a SMBH, as would be the case for those DWDs
forming in the Galactic Center, $\Gamma_{\rm acc}$ would be even greater. For
example, if we assume $m_3=4\times10^6\,M_\odot$ and $r=0.1$ pc, a DWD 
with ${\cal M}=0.3\,M_\odot$ and $f=2$ mHz will have $\Gamma_{\rm acc}=38$. The
apparent chirp mass will increase to ${\cal M}_o\simeq2.7\,M_\odot$. Now the
white dwarfs will masquerade as neutron stars or BHs!

So far, the estimation of the chirp mass is based on only two observables, $f$
and $\dot{f}$. One may wonder that given the long observational period of LISA
($\sim5$ years), could we derive also $\ddot{f}$ from the waveform? This
is an important question because if we could measure $\ddot{f}$, maybe we could
use it to break the degeneracy between mass and acceleration. In fact, for
BBHs, more advanced data analysis does show that it is possible
\cite{tamanini19}. 

\begin{figure}
\centering
\includegraphics[width=\textwidth]{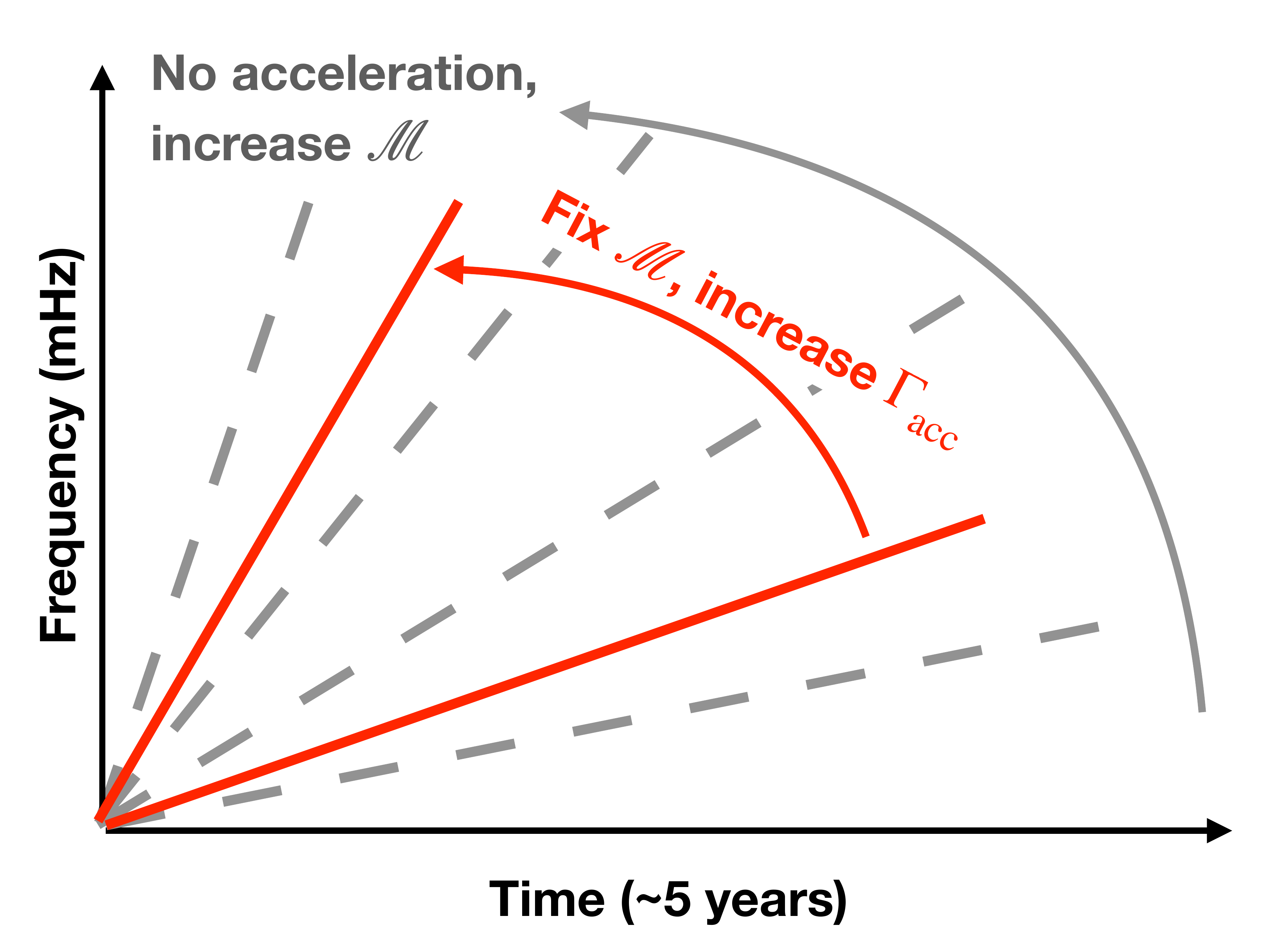}
\caption{Illustration of the degeneracy between mass and acceleration. The evolution
of the apparent frequency is effectively a straight line if the chirp mass of the source
is intrinsically small, such as in the case of DWDs. The gray dashed lines
show the dependence of the chirp signal on the chirp mass of the system. The red lines
show the dependence on the acceleration. It is clear that the effect of increasing
the acceleration is identical to the effect of increasing the chirp  mass.}
\label{fig:acc}
\end{figure}

For DWDs, however, it is more difficult
because the value of $\ddot{f}$ induced by either GW radiation or peculiar
acceleration is small. As Figure~(\ref{fig:acc}) shows, such a chirp signal is
featureless. The evolution of $f_o$ is effectively a straight line. Since we
use the slope of the straight line to infer the chirp mass, and small BBHs with
large acceleration lead to the same slopes as those big BBHs without acceleration,
it is difficult to distinguish these two scenarios \cite{robson18}.

\subsection{The effect of gas}

Gas is another important factor for the formation and evolution of BBHs.  
For example, in the model of binary-star evolution, there
is a phase when both binary members are immersed in a common gaseous envelope.
This phase is considered to be essential to the formation of close binaries.
Moreover, as has been mentioned before, AGN accretion disk is also a breeding
ground for BBHs. In general, gas will change the orbital dynamics of a BBH, and
in this way affect the GW signal. Without knowing of such an effect, would a
GW observer retrieve correctly the physical parameters of a BBH?

We now know that to address this question, we need to investigate the effect of
gas on $f$ and $\dot{f}$, and use them to derive the apparent chirp mass and
distance.  In fact, gas does not significantly affect $f$ because in the astrophysical
scenarios mentioned above, the total mass of the gas is negligible relative to
the total mass of the BHs. The effect is more prominent for $\dot{f}$.
Depending on the thermal dynamical properties and the distribution, gas could
either accelerate or decelerate the binary shrinkage, and hence increase or
decrease $\dot{f}$ relative to $\dot{f}_{\rm gw}$.  For example, hydrodynamical
friction\index{hydrodynamical friction}
makes the binary orbit shrink faster but tidal torque\index{tidal torque} 
and accretion could work in
the opposite direction (see a brief discussion in \cite{chen20gas}). 

\begin{figure}
\centering
\includegraphics[width=\textwidth]{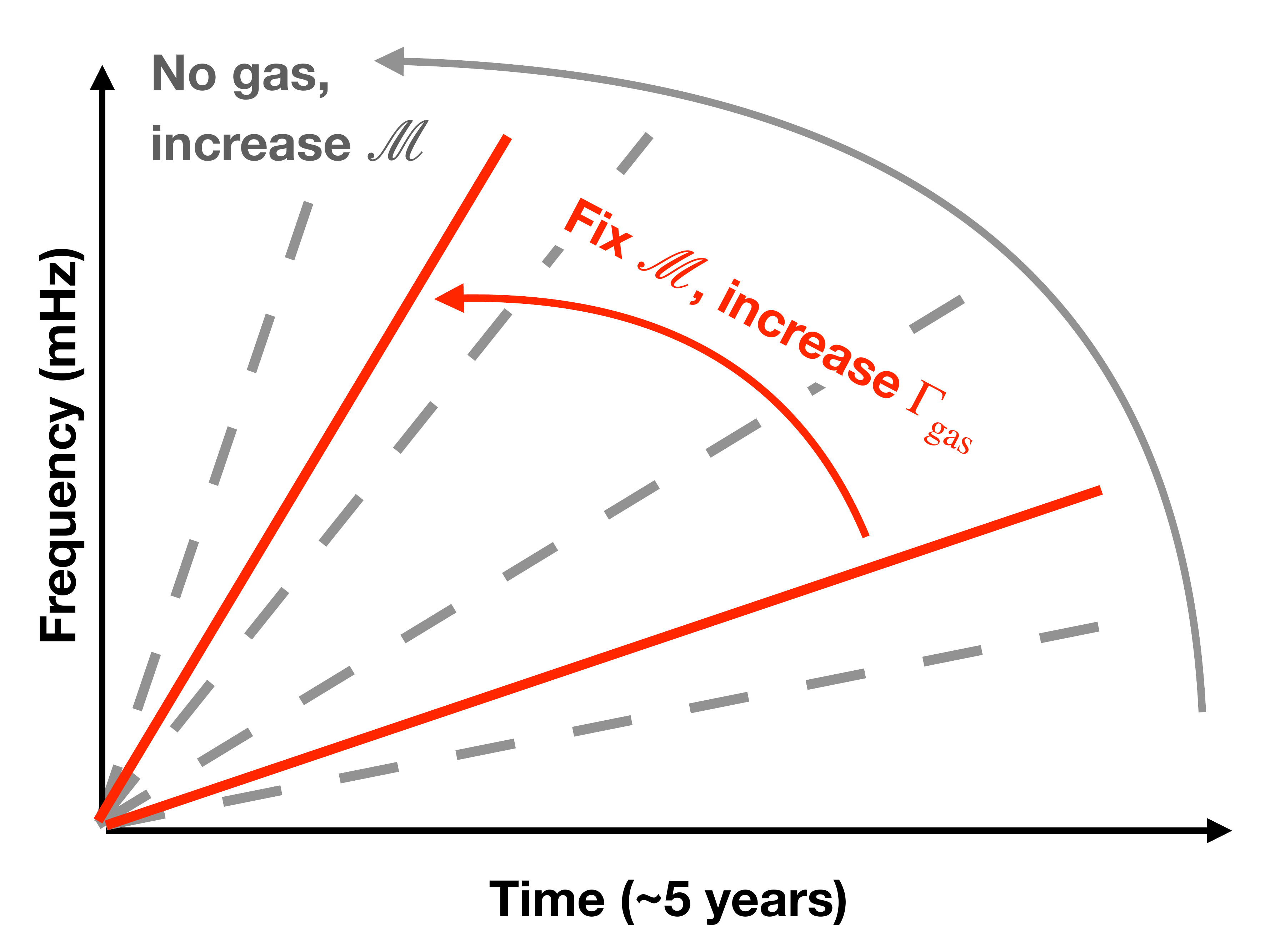}
\caption{Same as Figure~\ref{fig:acc} but showing the effect of gas on the
slope of the chirp signal.}
\label{fig:gas}
\end{figure}

To keep the following analysis general, we can characterize the gas effect by
rewriting the apparent chirp rate as
\begin{equation}
	\dot{f}_o=\dot{f}_{\rm gw}+\dot{f}_{\rm gas}=\dot{f}_{\rm gw}(1+\Gamma_{\rm gas}),
\end{equation}
where $\dot{f}_{\rm gas}$ is the chirp rate due to gas dynamics and
$\Gamma_{\rm gas}:=\dot{f}_{\rm gas}/\dot{f}_{\rm gw}$. 
The redshift effects are neglected here for simplicity.
Similar to the analysis of peculiar acceleration, we can derive that
\begin{align}
	{\cal M}_o&={\cal M}(1+\Gamma_{\rm gas})^{3/5},\\
	d_o&=d(1+\Gamma_{\rm gas}).
\end{align}
Since $\Gamma_{\rm gas}$ can be either positive or negative, the apparent mass
and distance could be bigger or smaller than their intrinsic values.
Figure~\ref{fig:gas} shows that the effect of increasing the efficiency of gas-dynamics (increasing
$\Gamma_{\rm gas}$) is identical to the effect of increasing the chirp mass.

Like the effect of peculiar acceleration, gas is more important for those
binaries in the mHz.  In this band, $\dot{f}_{\rm gw}$ is small so that the
gas-induced $\dot{f}_{\rm gas}$ could be relatively large.  Recent models show
that in AGN accretion disks and common envelopes, $\Gamma_{\rm gas}$ could be
much greater than $1$ \cite{chen19gas,caputo20,chen20gas}. Therefore, in these
environments BBHs, as well DNSs and DWDs, could appear more massive than they
really are.

\section{Summary}

In this chapter, we have seen that astrophysical factors could affect the
measurement of the masses of GW sources. In particular, strong gravitational
lensing, the Doppler and gravitational redshift around a SMBH, the peculiar
acceleration induced by a tertiary star or SMBH, and gas dynamics could all
increase the apparent chirp mass of a binary. In suitable conditions, the
Doppler blueshift in the vicinity of a SMBH, the peculiar acceleration, and gas
dynamics could also work in the opposite way, reducing the apparent mass.

These results have important implications for the peculiar compact objects
appearing in the bands of LIGO/Virgo and the future LISA. Such peculiar objects
include compact objects in the lower mass gap (between $3$ and $5\,M_\odot$)
and BHs heavier than $20\,M_\odot$.  They had never been detected in the past,
before the GW era. Therefore, their nature deserves a thorough scrutiny. 

\begin{table}
\caption{Alternatives models for the peculiar objects appearing in the LIGO/Virgo/LISA band.}
\label{tab:model}       
	\begin{tabular}{p{2cm}p{3cm}p{3.5cm}lc}
\svhline\noalign{\smallskip}
		Detector   & Peculiar Objects       & Imposters                    & Models & Notes  \\
\noalign{\smallskip}
\svhline\noalign{\smallskip}
		LIGO/Virgo & $\gtrsim20\,M_\odot$   & BBH $m_1,\,m_2<20\,M_\odot$  & 1, 2, 4  & i\\
			   &                        & BNS                          & --        & ii \\
			   &                        & DWD                          & --        & iii \\
		           & $(3,\,5)\,M_\odot$     & BBH $m_1,\,m_2\ge5\,M_\odot$ & 3        & iv\\
			   &                        & BNS                          & 2, 4     & v\\
			   &                        & DWD                          & --        & vi \\
\hline
		LISA       & $\gtrsim20\,M_\odot$   & BBH $m_1,\,m_2<20\,M_\odot$  & 6     & vii\\
			   &                        & BNS                          & 6        & viii \\
			   &                        & DWD                          & 6        & ix \\
		           & $(3,\,5)\,M_\odot$     & BBH $m_1,\,m_2\ge5\,M_\odot$ & 5, 6        & x \\
			   &                        & BNS                          & 5, 6     & xi \\
			   &                        & DWD                          & 5, 6        & xii\\
\noalign{\smallskip}
\svhline
\end{tabular}

	\footnotetext[1] 1 1. Gravitational lensing; 
	2. Doppler redshift (SMBH);
	3. Doppler blueshift (SMBH);\\ 
	4. Gravitational redshift (SMBH);
	5. Peculiar acceleration; 6. Gas dynamics.
\end{table}

Now we can revisit the question raised at the beginning of this chapter.  Once
LIGO/Virgo or LISA detect a peculiar object, could we readily accept the
apparent mass as the real one, or what are the other possibilities?
Table~\ref{tab:model} summarizes the alternative possibilities. The following
is a brief discussion of each case.  Interested readers could refer to the
previous sections for more detailed explanations. 

\begin{enumerate}[(i)]

\item In the LIGO/Virgo band ($10-10^2$ Hz), the BHs with an apparent mass higher than
$20\,M_\odot$ could be mimicked by the BHs less massive than $20\,M_\odot$, due
to strong lensing, or the Doppler and gravitational redshift around SMBHs. For the
BBHs in this band,  peculiar acceleration and gas dynamics are too weak
relative to GW radiation to cause such an strong effect.

	\item BNSs are unlikely imposters of the massive BBHs with
		$m_1,\,m_2>20\,M_\odot$ in the LIGO/Virgo band. In the
		scenarios of strong lensing, Doppler redshift, and
		gravitational redshift, the BNSs should have a redshift of
		$z\gtrsim10$ to meet the requirement. The corresponding
		magnification factor (${\cal A}$) would be too extreme, and the
		distance to SMBH too small. Such events are rare. Peculiar
		acceleration and gas dynamics are unlikely to take effect
		either.  Because GW radiation would predominate the evolution
		of $\dot{f}_o$ as soon as the BNSs enter the LIGO/Virgo band. 

	\item DWDs cannot masquerade as massive BBHs in the LIGO/Virgo band either. 
		Because of the
		large sizes of white dwarfs, the coalescence happens at a frequency of about
		$0.1$ Hz. Therefore, DWDs do not enter the LIGO/Virgo band.

	\item  The Doppler blueshift could also make an ordinary BH 
slightly more massive than $5\,M_\odot$
		appear in the lower mass gap in LIGO/Virgo observations.
		For example, a BBH at the ISCO of a Schwarzschild
		SMBH could be blueshifted. Combined with the gravitational
		redshift, the apparent mass could be a factor of $(1+z_{\rm
		dop})(1+z_{\rm gra})\simeq0.79$ of the real mass. For the IBO,
		the effect is more significant, and the fact could reduce to
		$0.58$. Around Kerr SMBHs, the factor could be even smaller. Peculiar
		acceleration and gas dynamics could hardly induce such an effect because,
		as is mentioned in (i), they are too weak compared to GW radiation. 

	\item BNSs merging in the vicinity of a SMBH could also appear in the lower-mass-gap
		in LIGO/Virgo observations, because of the Doppler and gravitational redshift.
		Strong lensing, on the other hand, is unlikely to cause such an effect, because
		BNSs have to be relatively nearby to be detected by LIGO/Virgo. Therefore, the probability
		of being lensed is low. For the same reason given in (ii), peculiar acceleration
		and gas dynamics are unlikely to move BNSs into the lower mass gap either.

	\item Same as (iii).  

	\item For LISA, the GW frequency is low. The corresponding
		BBHs are at an early evolutionary stage and the GW power is
		relatively weak. In this case, gas dynamics could compete with
		GW radiation and make less massive BH appear in the mass range
		of $\gtrsim20\,M_\odot$, while the effect of peculiar
		acceleration is still too weak to significantly affect $\dot{f}_o$.  Strong
		lensing is unlikely because LISA BBHs are at relatively small luminosity distances. 
		Doppler and gravitational redshift induced by a nearby
		SMBH is unlikely either, because LISA BBHs, due to their large
		semi-major axis, are likely to be tidally disrupted if they
		approach the Schwarzschild radius of a SMBH.

	\item BNSs could masquerade as BBHs in the LISA band if there is gas around. The other models
		are unlikely to work here for the same reasons given in (vii).

	\item The same as (vii) and (viii).

	\item Since gas dynamics could also reduce $\dot{f}_o$, it could make
		the BHs more massive than $5\,M_\odot$ appear in the low mass
		gap in LISA observations. Peculiar acceleration could work in
		the same way, but only when the BHs are slightly more massive than $5\,M_\odot$ and the tertiary is a SMBH.  Doppler blueshift around a SMBH could not
		induced such an effect because it would require BBHs to
		approach several Schwarzschild radii of the SMBH. LISA BBHs
		would be tidally disrupted at such a small distance. 

	\item BNSs could masquerade as lower-mass-gap objects in the LISA band due to the 
		peculiar acceleration and gas-dynamical effects. Strong gravitational lensing
		has difficulty inducing such an effect because LISA BNSs are relatively close by.
		Doppler and gravitational redshift due to a nearby SMBH is unlikely to work here,
		because of the strong tidal force close to the SMBH.

	\item The same as (xi), but the peculiar acceleration works only when the tertiary is a
		SMBH.

\end{enumerate}

Estimating the corresponding event rate could help us evaluate the likelihood
of the above models and, maybe, reject some of them
\cite{abbott19population,chen19}. But the estimation is subject to
uncertainties because of our incomplete understanding of the astrophysical
environments around GW sources. A more rigorous approach is to study the
distortion of the wave front by those astrophysical factors and look for a
detectable signature (e.g.
\cite{hannuksela19,torres-orjuela19,torres-orjuela20}).  Both directions
deserve further exploration. 


\bibliography{mybib}
\end{document}